# The basic open question of classical electrodynamics

Marijan Ribarič[1] and Luka Šušteršič[2]
Jožef Stefan Institute, p.p. 3000, 1001 Ljubljana, Slovenia

**ABSTRACT**

For the first time a method is devised for non-iterative modeling of motion of a radiating, electrified pointlike mass that has an internal structure. New, supplementary kinetic constants of accelerated charged particles are defined, that can be assessed by analysis of their trajectories in an accelerator.

[1] E-mail address: marjan.ribaric@masicom.net.

[2] Corresponding author. Tel.: +386 1 477 3258. Fax: +386 1 423 1569. E-mail address: luka.sustersic@ijs.si.

## 1. Introduction

In classical electrodynamics we study the implications of forces between electric charges and currents: and these forces affect their sources. About twenty years ago we published a monograph [1] about certain open questions of classical electrodynamics. Here we reconsider the basic open question of classical electrodynamics, since it turned out we should have followed the advice of English philosopher G.E. Moore in 1903, that to avoid difficulties and disagreements one should not "attempt to answer questions, without trying discovering precisely what question it is which you desire to answer," cf. Miller [2, Ch.1]. So let us point out that there are actually four, closely related but distinct subjects we are going to address:

a) In Section 2, we point out the basic conceptual question: Can classical electrodynamics provide at least one realistic dynamic model that takes a *non-iterative* account of the *radiation reaction force* (RRF), which is a direct consequence of the loss of four-momentum by the electromagnetic radiation.
b) In Sections 3 and 4, we consider such *differential four-momentum balance equations* for an *electrified pointlike mass* (EPM) that take account of the RRF – a subject of mathematical physics.
c) In Section 5, we mention the equation of motion for a *classical charged particle* – a subject of theoretical physics.
d) In Section 6, we propose a particular *experimental assessment* of the introduced *EPM's kinetic constants*, which take account of RRF's dynamic effects– a subject of experimental methodology.

## 2. The basic question of classical electrodynamics

Around 1602 Galileo began classical mechanics with the study of pendulums by an innovative *combination of experiment and mathematics*. Thereby he clearly stated that the laws of nature are mathematical.. However, if we only slightly rub a pendulum and electrify it, classical electrodynamics provides so far no non-iterative mathematical model of its swinging motion, which effects the loss of energy by electromagnetic radiation. As yet no detailed experimental observation of such a swinging motion has been performed, though physics is based on measurements.

According to Jackson [3], the basic trouble with classical electrodynamics is that we are able to obtain and study relevant solutions of its basic equations only in two limiting cases: "... one in which the sources of charges and currents are specified and the resulting electromagnetic fields are calculated, and the other in which external electromagnetic fields are specified and the motion of charged particles or currents is calculated… . Occasionally… the two problems are combined. But the treatment is a stepwise one -- first the motion of the charged particle in the external field is determined, neglecting the emission of radiation; then the radiation is calculated from the trajectory as a given source distribution. It is evident that this manner of handling problems in electrodynamics can be of only approximate validity." Thus there is no non-iterative modeling of interaction between electromagnetic fields and their sources. Consequently, we have: (a) only a partial physical understanding of such electromechanical systems where we cannot neglect this interaction, and (b) no evidence that the basic equations of classical electrodynamics are solvable.

## 3. A four-momentum balance equation for an electrified pointlike mass

In classical mechanics, the simple, idealized mathematical model of a pointlike mass is highly instructive and widely applicable, e.g. for studying the trajectories of planets. So it seems like it might be useful to have such a mathematical model about the motion of an electrified pointlike mass that takes account of RRF.

### *3.1. A differential four-momentum balance equation*

Inspired by the pointlike mass model, let us try to get some information about RRF's dynamic effects by considering the four-momentum of an accelerated EPM. Let the charge $q$ and mass $m$ of this EPM be located around the point $\mathbf{r}(t)$, and moving with velocity $\mathbf{v}(t)$ under the influence of a mechanical, Lorentz and/or gravitational external force $\mathbf{F}_{\text{ext}}(t)$; and let the relation between the four-force

$$f(t) = \gamma(\boldsymbol{\beta} \cdot \mathbf{F}_{\text{ext}}\,, \mathbf{F}_{\text{ext}})\,, \quad \text{where} \quad \gamma(t) = 1/\sqrt{1-|\boldsymbol{\beta}|^2} \quad \text{with} \quad \boldsymbol{\beta}(t) = \mathbf{v}/c\,, \tag{1}$$

the EPM's four-velocity $\beta(t) = (\gamma, \gamma\boldsymbol{\beta})$ be invariant under Poincaré transformations.[3]

Were $q = 0$, in classical mechanics we would model such a relation by the differential four–momentum balance equation

$$mc\beta^{(1)} = f \quad \text{with} \quad \beta^{(n)} \equiv (\gamma d\beta/dt)^n\,, \tag{2}$$

which is a heavily used, relativistic Newtonian equation of motion for a pointlike mass.

To take account of the RRF, we assume that EPM's electromagnetic radiation is adequately described by the Liénard-Wiechert potentials with singularity at $\mathbf{r}(t)$ , which emit the four-momentum

$$d\big(\beta^{(1)} \cdot \beta^{(1)}\big)\,\beta \quad \text{with} \quad d = q^2/6\pi\epsilon_0 c^2\,, \tag{3}$$

thereby diminishing EPM's four-momentum. Following Schott [4], we find that we must introduce the acceleration four-momentum $B(t)$ so as to obtain a complete description of changes to EPM's four-momentum effected by the RRF. Taking these two electromagnetic effects into account, we get from the equation of motion (2) of classical mechanics the following differential balance equation for EPM's four-momentum:

$$mc\beta^{(1)} - d\big(\beta^{(1)} \cdot \beta^{(1)}\big)\,\beta + B^{(1)} = f\,. \tag{4}$$

Dirac [5] concluded that the conservation of four-momentum requires that $B(t)$ is a four-vectored function of $f$ and $\beta$, and of a finite number of their derivatives, which satisfies the relation

---

[3] We will use the metric with signature $(+\,-\,-\,-)$, so that $\beta \cdot \beta = 1$.

$$\beta \cdot (B + d\beta^{(1)})^{(1)} \;=\; 0. \tag{5}$$

Thereafter, Bhabha [6] pointed out that the conservation of angular four-momentum requires that the cross product

$$\beta \wedge \left(B + d\beta^{(1)}\right) \tag{6}$$

is a total differential with respect to the proper time. Physical assumptions underlying the differential balance equation (4) are discussed in [1, Chs.9 and 10].

*3.2. Comments*

We cannot simplify EPM˙*s* balance equation (4) by disregarding the acceleration four-momentum $B$ carried by EPM˙*s* internal structure; because when $B$ = 0 and $q \neq 0$, then Dirac's and Bhabha's conditions (5) and (6) imply that $f$ = 0, i.e. there is no electrified EPM with $B$ = 0. However, we may pick $B = -d\beta^{(1)}$ so as to get apparently the simplest possible EPM˙*s* differential four-momentum balance equation:

$$mc\beta^{(1)} - \; d(1 - \beta\,\beta\,\cdot)\beta^{(2)} = \; f\;, \tag{7}$$

the so called Lorentz-Abraham-Dirac equation of motion for an electron. Such an equation of motion is considered questionable since (a) it is not a Newtonian kind of equation, and (b) it exhibits self-acceleration causing runaway solutions. The third term of the lhs (7) is known as the Abraham-Lorentz RRF; its relativistic modification, the sum of the second and third terms is called the Abraham-Lorentz-Dirac RRF. We will interpret this equation (7) in Section 7 as an asymptotic differential relation for EPM ˙s velocity.

Were certain acceleration four-momentum $B(t)$ given only as a function of EPM˙s velocity and of the external force, and satisfy Dirac's and Bhabha's conditions (5) and (6), the differential four-momentum balance equation (4) would be a relativistic Newtonian equation for the particular EPM specified by this $B(t)$. In [1, Secs.10.1 and 10.2] and [7], we pointed out seventeen qualitative properties that we are expecting from an equations of motion for a physically realistic EPM.

It is still an open question whether we can somehow augment the continuous classical electrodynamics with the notion of a pointlike charge, which is presently only a common and handy computational device. We generalized it by an expansion in terms of co-moving moments of time-dependent, moving charges and currents [8].

EPM seems to be the simplest generalization of classical mechanics notion of a pointlike mass to such a mathematical model of classical electrodynamics that takes into account RRF's dynamic effects. We considered it extensively in our monograph [1], naming it a classical pointlike charged particle. We found this name to be misleading, since we did not intend it to be synonymous with the concept of an elementary physical particle. So we now rename it "electrified pointlike mass".

*3.3. Electrically neutral pointlike mass with an internal structure*

EPM˙s balance equation (4) is also valid in the absence of electric charge, i.e. when $q = 0$. In such a case the differential four-momentum balance equation (4) reads

$$mc\beta^{(1)} + B^{(1)} = f. \quad (8)$$

Together with Dirac's and Bhabha's conditions that $\beta \cdot B^{(1)} = 0$ and that the cross product $\beta \wedge B$ is a total differential with respect to the proper time, the relation (8) models acceleration of a non-electrified pointlike mass with an internal structure specified by $B$. It is a generalization of the Newtonian equation of motion (2) and belongs to classical mechanics.

## 4. The equation of motion for a classical charged particle

In 1892, H. A. Lorentz started a century long quest to appropriately take account of RRF in modeling the motion of a classical charged particle, cf. [1, 9]. However, there are no pertinent quantitative experimental data. Recently Rohrlich [10] stated that, using physical arguments, he derived from the Lorentz-Abraham-Dirac equation (7) the physically correct equations of motion for a classical charged particle:

$$mc\beta^{(1)} = f + (d/mc)(1 - \beta\,\beta\,\cdot)f^{(1)}, \quad (9)$$

provided

$$\left|(d/mc)(1 - \beta\,\beta\,\cdot)f^{(1)}\right| \ll |f|. \quad (10)$$

As pointed out by Rohrlich [11], his equation (9) is a result of a *physical theory* and it is important to obey its validity limits (10) when testing it by experiments. This equation is not an EPM˙s model as specified by the relations (4) to (6).

## 5. An asymptotic differential relation for EPM˙s velocity in the case of a small and slowly changing external force

We do not know of such an acceleration four-momentum $B(t)$ that results in a physically realistic EPM ˙s differential four-momentum balance equation (4). So it is an open question whether there is one, none or many such balance equations that take account of RRF.

However, Dirac [5], Bhabha [6], and Eliezer [12] pointed out that in addition to $B = -d\beta^{(1)}$, there are also higher order relativistic polynomials in $\beta^{(n)}$ which satisfy Dirac's and Bhabha's conditions (5) and (6). So what would be the significance of a generalized Lorentz-Abraham-Dirac equation (7), were we to assume that EPM ˙s kinetic properties are modeled by a B(t) that is a sum of such polynomials? In 1989, inspired by the expansion of convolution-integrals in terms of derivatives of Dirac's delta function, we proposed that the four-momentum balance equation (4) with a particular combination of such polynomials provides an asymptotic expansion of EPM˙s acceleration four-momentum $B$ in the case of a small and slowly changing external force [13].

*5.1. An asymptotic, relativistic differential relation for EPM˙s velocity*

We put forward arguments [14; and 1, Chs.9-11] in support of the following hypothesis: Let the acceleration four-momentum $B(t)$ depend in a causal way, just on the values of acceleration $d\boldsymbol{v}(t)/dt$ from the time $t = 0$ on. And let the external force depend on a non-negative parameter $\lambda$ in such a way that $\mathbf{F}_{\text{ext}}(t) = \lambda \mathbf{F}(\lambda t)$, with $\mathbf{F}(t)$ being an analytic function of $t > 0$ and $\mathbf{F}(t \leq 0) = 0$. Then, in the asymptote $t \nearrow \infty$, the $n$th derivative $d^n \boldsymbol{v}(t)/dt^n$ of EPM ˙s velocity is of the order $\lambda^n$ as $\lambda \searrow 0$; and we may approximate the acceleration four-momentum $B(t)$ up to the order of $\lambda^5$ inclusive so that EPM ˙s velocity satisfies in the asymptote $t \nearrow \infty$ , up to the order of $\lambda^6$ inclusive, the following differential four-momentum balance equation:

$$\begin{aligned} mc\beta^{(1)} &- d(1 - \beta\,\beta\,\cdot)\beta^{(2)} \\ &+ e_1\left[\beta^{(2)} - \left(\beta \cdot \beta^{(2)} - {}^1\!/_2\, \beta^{(1)} \cdot \beta^{(1)}\right)\beta\right]^{(1)} \\ &+ e_2\left[\beta^{(4)} - \left(\beta \cdot \beta^{(4)} - \beta^{(1)} \cdot \beta^{(3)} + {}^1\!/_2\, \beta^{(2)} \cdot \beta^{(2)}\right)\beta\right]^{(1)} \\ &+ b_1\left[\left(\beta^{(1)} \cdot \beta^{(1)}\right)\beta^{(2)} + 2\left(\beta^{(1)} \cdot \beta^{(2)}\right)\beta^{(1)} + {}^7\!/_4 \left(\beta^{(1)} \cdot \beta^{(1)}\right)^2 \beta\right]^{(1)} = f, \end{aligned} \qquad (11)$$

where $d, e_1, e_2$ and $b_1$ are real parameters. The first term of the relativistic differential relation (11) is due to Einstein; the second one is due to Dirac who calculated that for an electron $d = e^2/6\pi\epsilon_0 c^2$ [5]; the relativistic polynomials in $\beta^{(n)}$ multiplied by $e_1$ and $e_2$ were constructed by Eliezer [12], and that multiplied by $b_1$ is due to Bhabha [6]. So let us refer to $d$ as the Dirac kinetic constant, to $e_1$ and $e_2$ as the Eliezer kinetic constants, and to $b_1$ as the Bhabha kinetic constant.

*5.2. Applications to calculations of the external forces and trajectories*

When the kinetic constants $m, d, e_1, e_2$ and $b_1$ of a particular EPM are known, one may use the differential relation (11), cf. [13] and [1, Sects.11.4 and 11.5]:

a) To determine up to the order of $\lambda^6$ inclusive the external force that effects certain EPM ˙s trajectory.
b) To modify (11) into a variety of approximate Newtonian equations of motion, such as the Rohrlich equation (9), by eliminating through iteration all higher order derivatives of velocity. And use them to calculate approximate EPM˙s trajectories.
c) To complete an ansatz, which we use to describe an asymptotic EPM ˙s trajectory, by expressing its parameters in terms of the external force constants and EPM ˙s kinetic ones.
d) To use it in construction of an empirical formula for the prediction of experimental data about EPM˙s trajectories, e.g. of an electrified pendulum, cf. [14].

## 6. Discussion

### *6.1. Experimental assessment of RRF's dynamic effects*

The Dirac, Eliezer and Bhabha parameters of the asymptotic differential relations (11) are supplementary kinetic constants, additional to the mass of a pointlike physical object, possibly electrified. We could assess them by fitting the observed trajectories corresponding to various external forces. To this end we might modify the asymptotic differential relation (11) by replacing the time derivatives with finite difference approximations, choosing step sizes consistent with the experimental data to be fitted. We have no quantitative suggestions about when we may expect to obtain in this way the consistent values of Dirac, Eliezer or Bhabha kinetic constants of an electrified physical object. So let us point out three qualitative conditions:

a) The object should be pointlike in the sense of Pauli [15, §29]; so were its charge $q$ negligible, then its equation of motion should be an equation of motion for a pointlike mass of classical mechanics.
b) It should be losing the four-momentum through the radiation as effected by the Liénard-Wiechert potentials.
c) The external force should be small and slowly changing as specified above.

Inspired by J. J. Thomson, who could made good estimates of both the charge and mass of electron in 1898 by observing its trajectories, we could seek to assess the Dirac, Eliezer and Bhabha kinetic constants of a particular accelerated physical particle by using the differential relation (11) in fitting its velocities. To observe them we could use particle accelerators, since they may accelerate particles of any mass, from electrons and positrons to uranium ions. Such an innovative use of accelerators, for considering also classical kinetic properties of physical particles, may provide new insight into particle acceleration, helpful for improvement of performance and design of accelerator facilities.

In 1797-98, H. Cavendish performed the first laboratory experiments to measure the force of gravity between masses. This research is ongoing [16], and may inspire the experiments to measure the Dirac, Eliezer and Bhabha kinetic constants of an electrified pointlike mass by using the differential relation (11).

### *6.2. Testing theories*

The potentially experimentally testable results of the theories that take account of RRF are the proposed theoretical equations of motion for a classical charged particle. Each equation of motion implies specific asymptotic expansion of velocity analogous to (11), which we can then test experimentally. E.g. up to the order of $\lambda^3$ inclusive, such an asymptotic differential relation corresponding to the Rohrlich equation (9) reads:

$$mc\beta^{(1)} - d(1-\beta\,\beta\,\cdot)\beta^{(2)} + \left(d^2/mc\right)\left[\beta^{(3)} - \beta\cdot\beta^{(3)}\,\beta + \beta^{(1)}\cdot\beta^{(1)}\,\beta^{(1)}\right] = f\,. \qquad (12)$$

Considering the kinetic properties of an electron (or positron), one might check Dirac's assumption [5], which he made when deriving Lorentz-Abraham-Dirac equation (7), that an electron is such a simple thing that $B = -d\beta^{(1)}$; thus according to Dirac, the Eliezer and Bhabha kinetic constants of an electron should turn out to be negligible.

## 7. Conclusions

By avoiding iterations, the asymptotic differential relation (11) for EPM ˙s velocity provides the first non-iterative modeling of RRF's dynamic consequences. It defines new, supplementary kinetic constants of an accelerated electrified pointlike mass, which can be assessed by observing their trajectories in an accelerator. And it suggests that the basic equations of classical electrodynamics are solvable.

As the Lorentz-Abraham-Dirac equation (7) equals the asymptotic differential relation (11) with $e_1 = e_2 = b_1 = 0$, it makes sense to interpret it as the first differential relation for EPM ˙s velocity that that takes some account of the RRF in the case of a small and slowly changing external force.